%% file: r2017.tex
\documentclass[10pt]{article}
\usepackage{graphicx}

\def\Title#1{\begin{center} {\Large #1 } \end{center}}
\def\Author#1{\begin{center}{ \sc #1} \end{center}}
\def\Address#1{\begin{center}{ \it #1} \end{center}}

\newcommand\pubblock{\rightline{\begin{tabular}{l} Proceedings of the Fifth Annual LHCP\\ \pubnumber\\
         \pubdate  \end{tabular}}}

\newenvironment{Abstract}{\begin{quotation} \begin{center} 
             \large ABSTRACT \end{center}\bigskip 
      \begin{center}\begin{large}}{\end{large}\end{center} \end{quotation}}

\newenvironment{Presented}{\begin{quotation} \begin{center} 
             PRESENTED AT\end{center}\bigskip 
      \begin{center}\begin{large}}{\end{large}\end{center} \end{quotation}}

\def\Acknowledgements{\bigskip  \bigskip \begin{center} \begin{large}
             \bf ACKNOWLEDGEMENTS \end{large}\end{center}}

\input econfmacros.tex

\usepackage{color}

\newcommand\pubnumber{CERN-TH-2017-223\\LAPTH-Conf-035/17}

\newcommand\pubdate{\today}

\def\affiliation{
CERN, Theoretical Physics Department, Geneva, Switzerland\\
LAPTh, CNRS, Universit\'e Savoie Mont Blanc, Annecy-le-Vieux, France}

\def\support{\footnote{This research project has been supported by a Maria Sk\l{}odowska-Curie Individual Fellowship of the European Commission's Horizon 2020 Programme under contract number 659147 PrecisionTools4LHC.}}

\newcommand{\pth}{p_{T,H}}
\newcommand{\ptveto}{p_{T,veto}}

\begin{document}

\large
\begin{titlepage}
\pubblock

\vfill
\Title{ Predictions for exclusive Higgs cross sections  }
\vfill

\Author{ Emanuele Re \support }
\Address{\affiliation}
\vfill
\begin{Abstract}

  I give an overview of recent theoretical results for exclusive
  Higgs-production cross sections, focusing in particular on processes
  where precise predictions will be relevant in the near future.

\end{Abstract}
\vfill

\begin{Presented}
The Fifth Annual Conference\\
 on Large Hadron Collider Physics \\
Shanghai Jiao Tong University, Shanghai, China\\ 
May 15-20, 2017
\end{Presented}
\vfill
\end{titlepage}
\def\thefootnote{\fnsymbol{footnote}}
\setcounter{footnote}{0}
%

\normalsize 


\section{Introduction}
Regardless of whether hints of new-Physics in direct searches will
show up in current and future LHC Runs, precision measurements in
Higgs studies will play a major role in the future. This is
particularly true because the increase in luminosity and collision
energy in Run II and beyond will allow for measurements of
differential distributions. In order to fully exploit the latter, it's
necessary that the theoretical accuracy for signal predictions matches
the experimental one. In this manuscript, I review some of the recent
developments relevant to this end. First I will focus on ``parton
level'' predictions, whereas, in the second part, I will review status
and recent news in the development of fully exclusive ``MC
generators''.

I will mostly focus on higher-order results in ``perturbative QCD'',
and only mention a few cases of recent progress in higher-order
perturbative computations in the ``EW'' couplings. Moreover, I will
not discuss recent results on the computation of higher-order
corrections for the Higgs-boson decays. For further details and a
fully comprehensive review, I advise the reader to refer to
ref.~\cite{deFlorian:2016spz}.

\section{Parton-level results}
\label{sec:parton}
Higgs production at the LHC is usually categorized according to its
four main production channels, namely ``gluon-fusion production''
(henceforth denoted as ``$gg\to H$''), ``vector-boson fusion
production'' (VBF), ``associated production'' ($VH$) and
``top-associated production'' ($t\bar{t}H$). The state of the art for
cross sections for these processes is summarized in
table~\ref{tab:table1}, where I specify the perturbative order at
which the fully inclusive result, as well as the order at which
differential distributions for ``Born-like'' observables
($\sigma_{tot}$ and $d\sigma$, respectively), are
known.\footnote{These are the observables whose definition does not
  involve any requirement on initial or final state radiation, except
  on the one already present at leading order, as in VBF. For example,
  the Higgs rapidity in gluon-fusion, or the transverse momentum of a
  tagging-jet in VBF, belong to this category.}
\begin{table}[h]
\begin{center}
  \begin{tabular}{l|c|c|c}
~ &  QCD order &  EW order &  other corrections known \\ \hline
$gg\to H$ &
$\sigma_{tot}$: N3LO, $d\sigma$: NNLO &
$\sigma_{tot}, d\sigma$: NLO &
N3LL threshold resummation \\
~ & ~ &  ~ & $m_q$ dependence: NLO (approx. NNLO) \\ 
~ & ~ &  ~ & mixed QCD/EW corrections \\ \hline
VBF &
$\sigma_{tot}$: N3LO (NNLO), $d\sigma$: NNLO &
$\sigma_{tot}, d\sigma$: NLO \\ \hline
$VH$ &
$\sigma_{tot}, d\sigma$: NNLO &
$\sigma_{tot}, d\sigma$: NLO &~\\ \hline
$t\bar{t}H$ &
$\sigma_{tot}, d\sigma$: NLO &
$\sigma_{tot}, d\sigma$: NLO &
NNLL threshold resummation ~\\ \hline
\end{tabular}
\caption{Total and differential cross sections for the four main Higgs
  production processes at hadron colliders.  Notice that the VBF total
  cross-section is known at N3LO~\cite{Dreyer:2016oyx} but only in the
  structure-function approach, where it's not possible to apply the
  standard cuts on the two leading jets which define the VBF signal
  region. The NNLO result of ref.~\cite{Cacciari:2015jma} is instead
  fully differential.}
\label{tab:table1}
\end{center}
\end{table}

Together with the above results, the knowledge of more
exclusive distributions will become more and more important in the
years to come: the transverse-momentum spectrum of the Higgs-boson, or
the cross section for producing the Higgs-boson in association with
jets (1, 2, as well as 0), are such an example. In the rest of this
section, I'll review some recent developments in this context,
focusing primarily on the gluon-fusion and VBF production mechanisms.

\subsection{The Higgs-boson transverse momentum}
\begin{figure}[htb]
\centering
\includegraphics[height=1.8in]{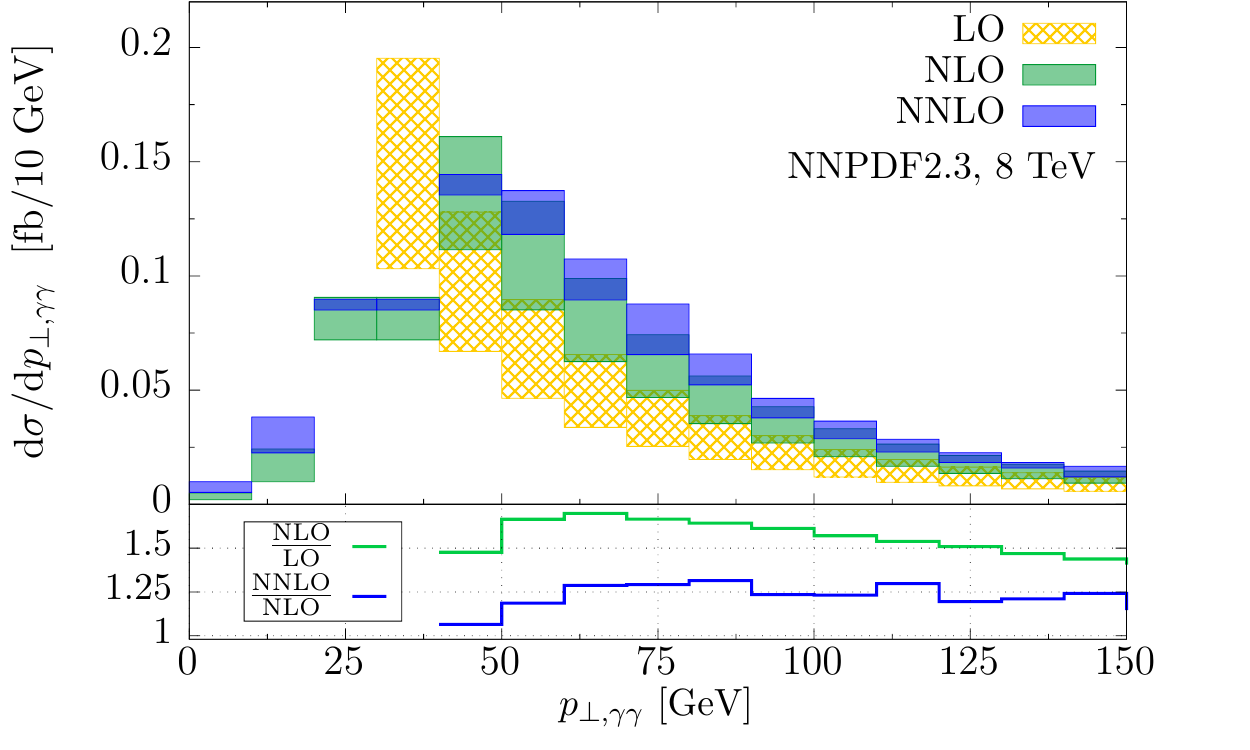}\hspace{0.5cm}
\includegraphics[height=1.8in]{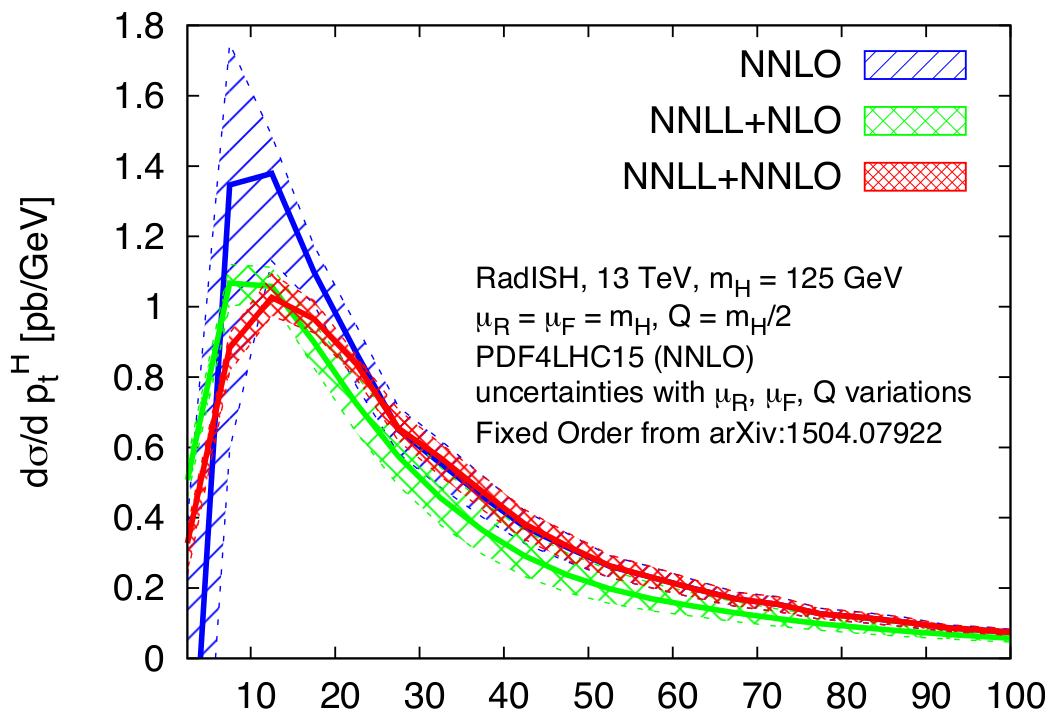}
\caption{Left: $\pth$ distribution at LO, NLO and NNLO, in presence of
  fiducial cuts on the photons from the Higgs decay, as well as with a
  cut on the hardest jet (plot taken from
  ref.~\cite{Caola:2015wna}). Right: resummed predictions up to
  NNLL+NNLO for $\pth$ as obtained with the {\tt Radish} code,
  developed in refs.~\cite{Monni:2016ktx,Bizon:2017rah}.}
\label{fig:pth}
\end{figure}

The Higgs transverse-momentum ($\pth$) spectrum in gluon fusion is now
known at NNLO in
QCD~\cite{Boughezal:2015dra,Boughezal:2015aha,Chen:2016zka}, although
exact corrections beyond leading order have been computed only in the
HEFT approach,~\emph{i.e.} in the limit $m_t\to\infty$.  As shown in
the left-hand plot in fig.~\ref{fig:pth}, the NNLO/NLO K-factor is large and non
flat: including NNLO corrections not only reduces the theoretical
uncertainties, but also improves the data-theory agreement.

At small values of $\pth$ (below $\sim 30$ GeV) large logarithms
of the type $\log(m_H/\pth)$ arise at all orders, due to (multiple)
soft/collinear radiation, and in order to obtain a meaningful and
well-behaved prediction, an all-order resummation is needed.
Until recently the more accurate results were those obtained in
refs.~\cite{Bozzi:2003jy,Becher:2012yn}, where the resummation was
carried out at the NNLL+NLO order,~\emph{i.e.}~the NNLL resummed
result was matched to the NLO computation of the differential cross
section $d\sigma/d\pth$, yielding also NNLO accuracy for the
integrated spectrum. In two recent
papers~\cite{Monni:2016ktx,Bizon:2017rah}, a new method was formulated
to resum small $\pth$ logarithms directly in momentum space, avoiding
at the same time the spurious singularities in the spectrum usually
arising when resumming in direct space. The method was used to obtain
NNLL+NNLO accurate results, and, more recently, to achieve N3LL+NNLO
accuracy~\cite{Bizon:2017rah}.\footnote{The NNLO $\pth$ spectrum and
  the total N3LO cross sections used in
  refs.~\cite{Monni:2016ktx,Bizon:2017rah} were taken from
  refs.~\cite{Boughezal:2015dra}
  and~\cite{Anastasiou:2015ema,Anastasiou:2016cez}, respectively.} An
example of the numerical relevance of some of these effects is shown
in the right-hand side panel of fig.~\ref{fig:pth}, which has been
obtained using the {\tt Radish} code. Although numerical results have
not been published yet, a different approach to perform this
resummation in direct space was also formulated in
ref.~\cite{Ebert:2016gcn}.

Hints of new-Physics can be found by looking at the Higgs transverse
momentum spectrum. For instance, it is known that in the large $\pth$
limit of gluon-fusion production (boosted Higgs regime), effects due
to the presence of new heavy states circulating in the loop can be
exposed, since a boosted Higgs boson probes the heavy degrees of
freedom in virtual corrections. Although this idea is very powerful, it
is clear that a very precise knowledge of the SM result in the boosted
regime is needed in order to fully exploit it.  Of chief importance
are the top-mass effects beyond leading order: it is indeed known that
the heavy-top approximation significantly overestimates the exact LO
result, as first observed in refs.~\cite{Ellis:1987xu,Baur:1989cm}.
Currently these corrections are not known, because some 2-loop
integrals have not been fully computed yet~\cite{Bonciani:2016qxi}. On
the other hand, one loop amplitudes with full top-mass dependence are
known for $H+2$~\cite{DelDuca:2001fn} and
$H+3$~\cite{Campanario:2013mga,Greiner:2016awe} jets. By using them,
different strategies to estimate the complete result have been put
forward: for instance, improving upon previous
results~\cite{Harlander:2012hf}, the authors of
ref.~\cite{Neumann:2016dny} used the exact 1-loop results for $H+1$
and $H+2$ partons matrix elements together with an expansion up to
$\mathcal{O}(m_t^{-4})$ to estimate the (unknown) finite parts of the
massive virtual (2-loop) corrections. A reweighting procedure was used by
the authors of ref.~\cite{Chen:2016zka} to include the unknown mass
effects in their NNLO computation. The authors of
ref.~\cite{Braaten:2017lxx} have instead proposed a method to perform
a systematic approach in powers of $m_t/\pth$. In
ref.~\cite{Caola:2016upw} the high-energy behavior of
the Higgs plus jet have also been used to estimate mass effects in
this regime. Other results important for current phenomenology have
been obtained in the context of NLO+PS merging, as discussed in
sec.~\ref{sec:MC}.

\begin{figure}[htb]
\centering
\includegraphics[height=2.2in]{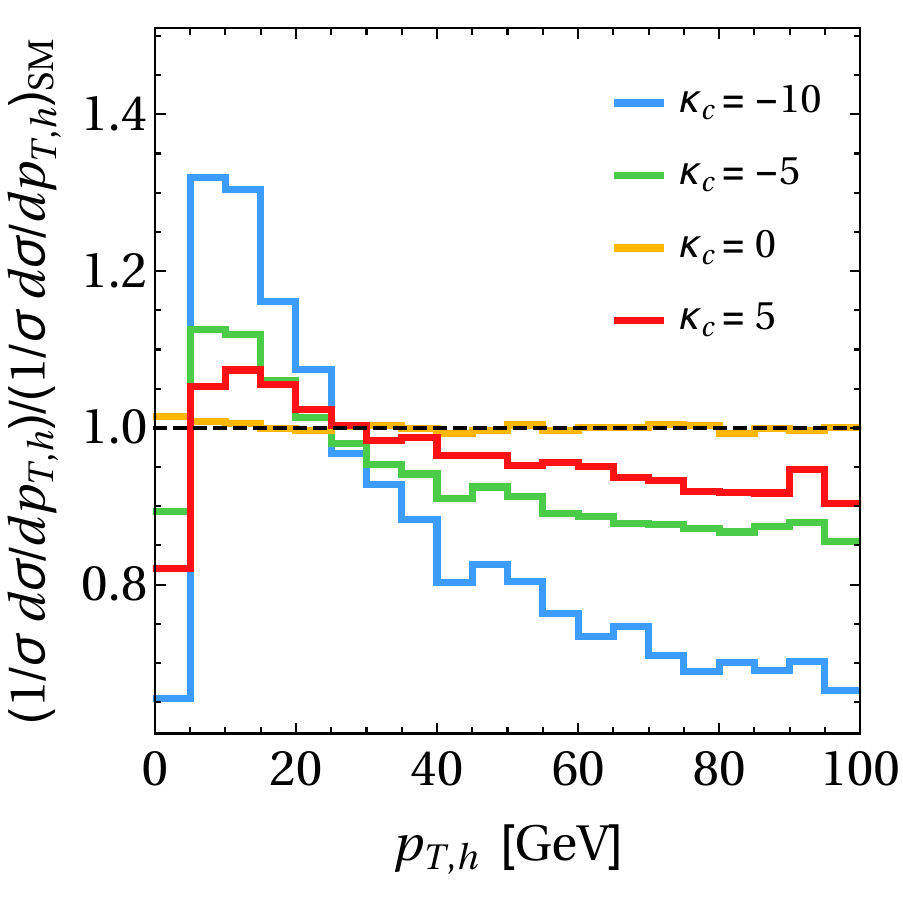}\hspace{1cm}
\includegraphics[height=2.2in]{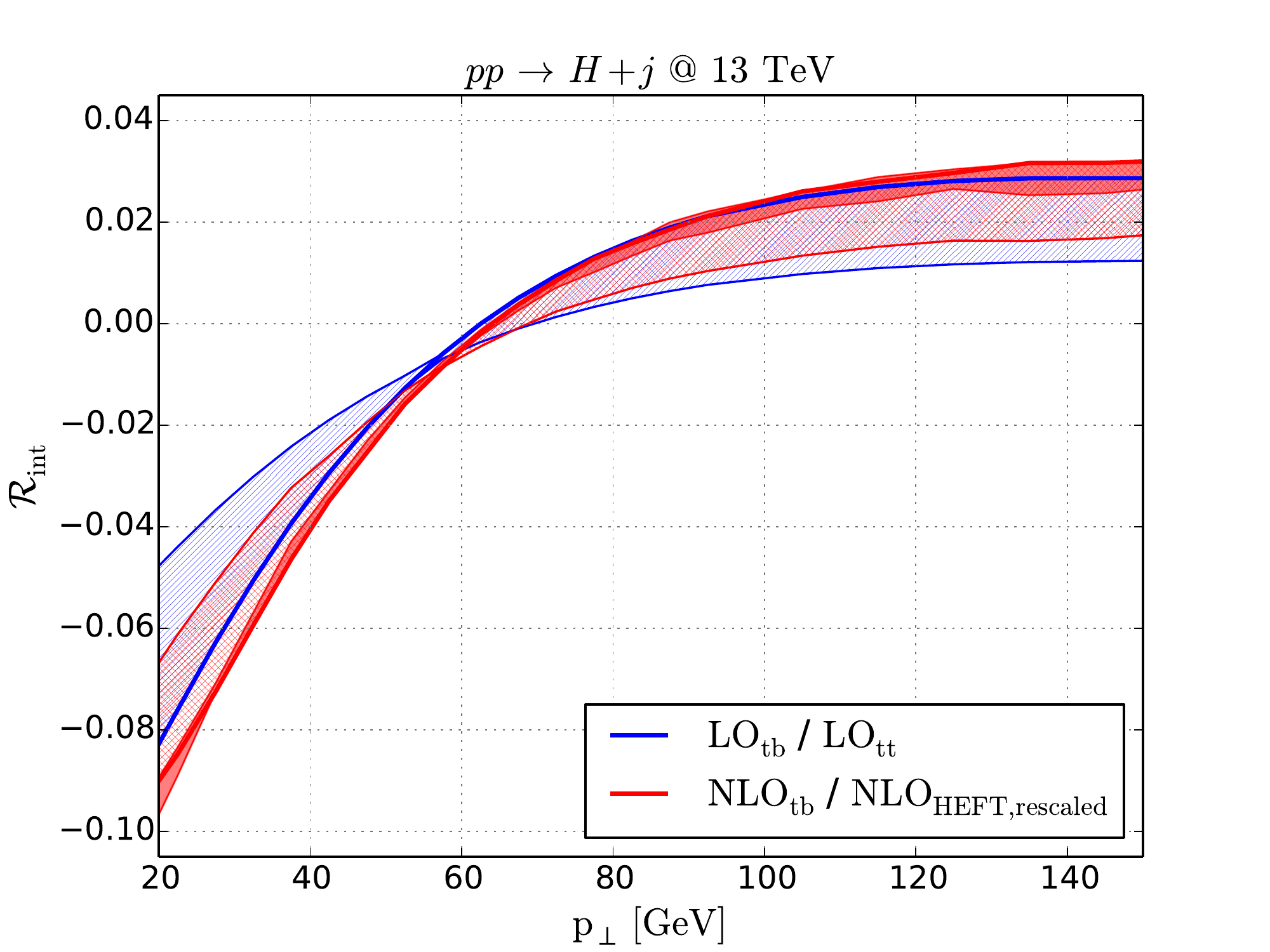}
\caption{Left: normalised $\pth$ spectrum of inclusive Higgs
  production divided by the SM prediction for different values of
  $\kappa_c$ (plot taken from ref.~\cite{Bishara:2016jga}). Right:
  relative top-bottom interference contribution to the transverse
  momentum distribution of the Higgs boson at LO and NLO (plot taken
  from ref.~\cite{Lindert:2017pky}).}
\label{fig:bcmass}
\end{figure}

Important properties of the Higgs boson can also be inferred by
looking at the medium-small $\pth$ region. For instance, a method to
set competitive constraints on the bottom and charm Yukawa couplings
$y_Q=\kappa_Q\ y_{Q,SM}$ ($Q=b,c$) was proposed in
ref.~\cite{Bishara:2016jga}.\footnote{Similar ideas were also
  suggested in ref.~\cite{Soreq:2016rae}.} The main idea is based on
the following observation: among the Higgs production mechanisms, two
of them depend on the charm (and bottom) Yukawa coupling, and are, at
the same time, mostly relevant at medium-small $\pth$,~\emph{i.e.}
gluon fusion ($gg\to Hg$) and heavy-quark-initiated ($gQ\to HQ$,
$Q\bar{Q}\to Hg$) production. The cross section for the latter case
depends upon $y_Q^2$, whereas, for gluon-fusion, the leading dependence
is linear in $y_Q$, as it originates from the interference among $2\to
2$ diagrams with a top loop and those with a bottom/charm loop,
schematically $d\sigma^{LO}_{tQ}\sim \Re \{A_t^{1\mbox{-}loop} A_Q^{*,1\mbox{-}loop}
\}$. Moreover, in the gluon fusion case, and in the regime
$m_Q\ll \pth \ll m_H$, such interference features also (non-Sudakov)
double logarithms of the form $\kappa_Q (m_Q/m_H)^2
\log^2{(m_Q^2/\pth^2)}$. As a consequence, the final differential
cross-section $d\sigma/d\pth$ (as well as the leading-jet
distribution) is quite sensitive to variations of the bottom and charm
Yukawa modifiers $\kappa_Q$, particularly so at medium $\pth$, as
shown for instance in the left panel of fig.~\ref{fig:bcmass}. Thanks to this
dependence, by using Run I data, and uncertainties thereof, together
with the currently available theory predictions (where the total
uncertainty amounts to 10\% at most), the constraint $\kappa_c\in
[-16,18]$ at 95\% CL was obtained in ref.~\cite{Bishara:2016jga}. If
one assumes a total experimental uncertainty (dominated by systematic)
of 3\%, and a total theory uncertainty of 5\%, it was shown
in~\cite{Bishara:2016jga} that, with $300\ \mbox{fb}^{-1}$, the
projection for the constraint on $y_c$ reads $\kappa_c\in [-1.4,3.8]$
at 95\% CL, which is at least as good as what can be obtained with all
the other available
methods~\cite{Bodwin:2013gca,Kagan:2014ila,Koenig:2015pha,Perez:2015lra,Perez:2015aoa,Brivio:2015fxa}
taken individually.

It is clear that, in the the long term, the bottleneck to fully
exploit this idea will be the quality of the theory prediction. The
most notable missing ingredient is the knowledge of the NLO
corrections for $d\sigma/\pth$ with exact mass effects. For the case
at hand, this reduces to the knownledge of $d\sigma^{virt}_{tQ}\sim
\Re\{A_t^{1\mbox{-}loop} A_Q^{*,2\mbox{-}loop} + A_t^{2\mbox{-}loop}
A_Q^{*,1\mbox{-}loop}\}$,~\emph{i.e.} the virtual correction to
$d\sigma^{LO}_{tQ}$.  In this respect, the results in
refs.~\cite{Lindert:2017pky} are
extremely relevant: for the bottom-quark case, the authors of
refs.~\cite{Melnikov:2016qoc,Melnikov:2017pgf} computed $A_b^{*,2
  loop}$ by consistently neglecting all the terms that are
power-suppressed in the $m_b\to 0$ limit, whilst keeping all the
non-analytic $\mathcal{O}(\log(m_b))$ terms. This is exactly the
approximation needed to study the $m_b$ effects on the $\pth$ spectrum
in the intermediate range. By combining this result with the exact
real-emission matrix elements, as well as using $A_t^{2\mbox{-}loop}$
in the $m_t\to\infty$ limit (which is perfectly adequate for $\pth\ll
m_t$), a complete NLO result for $d\sigma_{tb}$ was published in
ref.~\cite{Lindert:2017pky}: these corrections were found to be
important($\mathcal{O}(40\%)$) and similar to the NLO corrections for
$d\sigma_{tt}$, both in size and shape. Moreover, renormalization
scheme ambiguities on the treatment of $m_b$ were shown to reduce at
NLO, at least for $\pth<60$ GeV (see for instance the right panel in
fig.~\ref{fig:bcmass}). It will be extremely interesting to combine
these results with the state-of-the-art computations of $\pth$
mentioned in the first part of this section. Moreover, the results of
ref.~\cite{Lindert:2017pky} will allow to quantify more precisely the
ultimate reach of the idea presented in ref.~\cite{Bishara:2016jga} to
set bounds on the light-quark Yukawa couplings.

\subsection{Cross section in jet bins}
In several analyses relevant for precision Higgs measurements, events
are categorized according to the number of jets accompanying the
Higgs-boson decay products. An example is the $0$-jet cross section
(often called ``jet-vetoed'' cross section), where a jet-veto $\ptveto
\approx 30$ GeV is required for the hardest jet, in order to suppress
the $t\bar{t}$ backgrounds when looking for $H\to WW$ and $H\to
\tau\tau$ in gluon-fusion dominated searches. Another obvious example
is the 2-jet cross section, which will become more and more important
for VBF studies. In the rest of this section, I'll summarize recent
progress made for the signal prediction in these two cases.

When a jet-veto is applied to the $gg\to H$ process, logarithms of the
type $\log(m_H/\ptveto)$ are generated. Although a fixed order
computation is known to yield a very accurate result for the central
value of the jet-vetoed cross section, an all-order resummation is
necessary if a perturbative uncertainty of few percent on the signal
rate is sought for.\footnote{A
  thourough discussion of the theoretical predictions for the
  backgrounds is equally important, but it goes beyond the scope of
  this manuscript.} The more advanced results for this quantity were
published in ref.~\cite{Banfi:2015pju}, where the NNLL resummed
results obtained by the authors of~\cite{Banfi:2012jm} were matched to the
fully-inclusive N3LO cross section~\cite{Anastasiou:2015ema,Anastasiou:2016cez} and the H+1 jet NNLO
spectrum~\cite{Boughezal:2015dra}. Mass effects were also taken into account following
the scheme discussed in~\cite{Banfi:2013eda} and a leading-log resummation of
jet-radii logarithms was also included~\cite{Dasgupta:2014yra}. The final
perturbative uncertainty for $\ptveto\sim 30$ GeV amounts to about
4\%: about half of that is due to the N3LO result, and the residual
part comes from the resummation accuracy.

In Run II and beyond, the measurements of VBF production will be
particularly important. For instance, VBF Higgs production is the
largest production mode involving only tree-level interactions, and,
since $\pth$ is already non-vanishing at the leading order, it
facilitates searches of Higgs invisible decay modes. One of the major
background components for the extraction of the VBF signal
is the gluon-fusion production of Higgs accompanied by two (or more)
jets. In order to ennhance the signal over the background, the
kinematics of the final state objects is heavily used. Although NNLO
corrections to the total cross-section within VBF cuts are extremely
small ($\mathcal{O}(1\mbox{-}2\%)$), in ref.~\cite{Cacciari:2015jma}
it was found that they are not flat across the fiducial phase space,
as they can reach up to $\mathcal{O}(10\%)$. Not only these results
are extremely important on their own, but they also allow to expose
current limitations of NLO+PS tools: in fact, although at times
higher-order effects are (approximately) captured by these tools, in
ref.~\cite{Cacciari:2015jma} it was also found that, for VBF, this is
not necessarily the case for all observables.  As currently one of the
largest theory uncertainties for VBF searches comes from NLO+PS
generators~\cite{ATLAS:2016gld}, an interesting avenue for future
developments will be to find an optimal way to use the NNLO results to
improve MC tools. Eventually, it will be desirable to match these NNLO
results to parton-showers, as it has been already done for $gg\to H$
and $VH$ production (see sec.~\ref{sec:MC}).

\section{Monte Carlo event generators}
\label{sec:MC}
Nowadays NLO+PS Monte Carlo event generators are well established and automated,
and I will not attempt to reference all the available tools and
implementations: table~\ref{tab:table2}, though, summarizes
schematically the current state of the art.
\begin{table}[h]
  \begin{center}
\begin{tabular}{l|c|c}  
~ &  matching &  NLO+PS multi-jet merging \\ \hline
$gg\to H$ & NNLO+PS & $d\sigma_{H+0j} + d\sigma_{H+1j} + d\sigma_{H+2j}$ \\ \hline
VBF & NLO+PS ($d\sigma_{Hjj}$, $d\sigma_{Hjjj}$) & - \\ \hline
$VH$ & NNLO+PS & $d\sigma_{VH+0j} + d\sigma_{VH+1j}$ \\ \hline
$t\bar{t}H$ & NLO+PS & -\\ \hline
$pp\to HH$ & NLO+PS (exact $m_t$ dependence) & -\\ \hline
\end{tabular}
\caption{Summary of available fully-exclusive Monte-Carlo generators
  for the simulation of Higgs signal cross sections.}
\label{tab:table2}
\end{center}
\end{table}

As far as relatively recent NLO+PS results are concerned, in
ref.~\cite{Jadach:2016qti} $gg\to H$ production was matched to parton
showers using the so-called {\tt KrKNLO} method~\cite{Jadach:2015mza},
which is an approach different with respect to the commonly used {\tt
  MC@NLO} and {\tt POWHEG} methods. Another important development is
that not only has the differential cross section for di-Higgs
production been computed at NLO, including the exact dependence upon
$m_t$~\cite{Borowka:2016ehy,Borowka:2016ypz}, but also that this
computation, which is numerically very heavy due to the presence of
massive 2-loop amplitudes, has been matched to parton
shower~\cite{Heinrich:2017kxx} using the {\tt POWHEG} method.

Aside from the aforementioned recent NLO+PS results, most of the more
recent activity has focussed on the concept of NLO+PS multijet
merging, which collectively denotes all the techniques aiming at
describing with NLO+PS accuracy several jet multiplicities (possibly
accompanied by a massive system) in a single event sample. To achieve
this goal, several ideas have been put forward: {\tt
  MEPS@NLO}~\cite{Hoeche:2012yf,Gehrmann:2012yg}, ``{\tt FxFx}''
merging~\cite{Frederix:2012ps}, {\tt
  UNLOPS}~\cite{Lonnblad:2012ix,Platzer:2012bs,Bellm:2017ktr}, {\tt
  MiNLO}~\cite{Hamilton:2012np,Hamilton:2012rf}, {\tt
  Geneva}~\cite{Alioli:2012fc,Alioli:2013hqa}, {\tt
  Vincia}~\cite{Hartgring:2013jma,Fischer:2016vfv}. The former four
have been already used for processes relevant for Higgs Physics.

\begin{figure}[htb]
\centering
\includegraphics[height=2.0in]{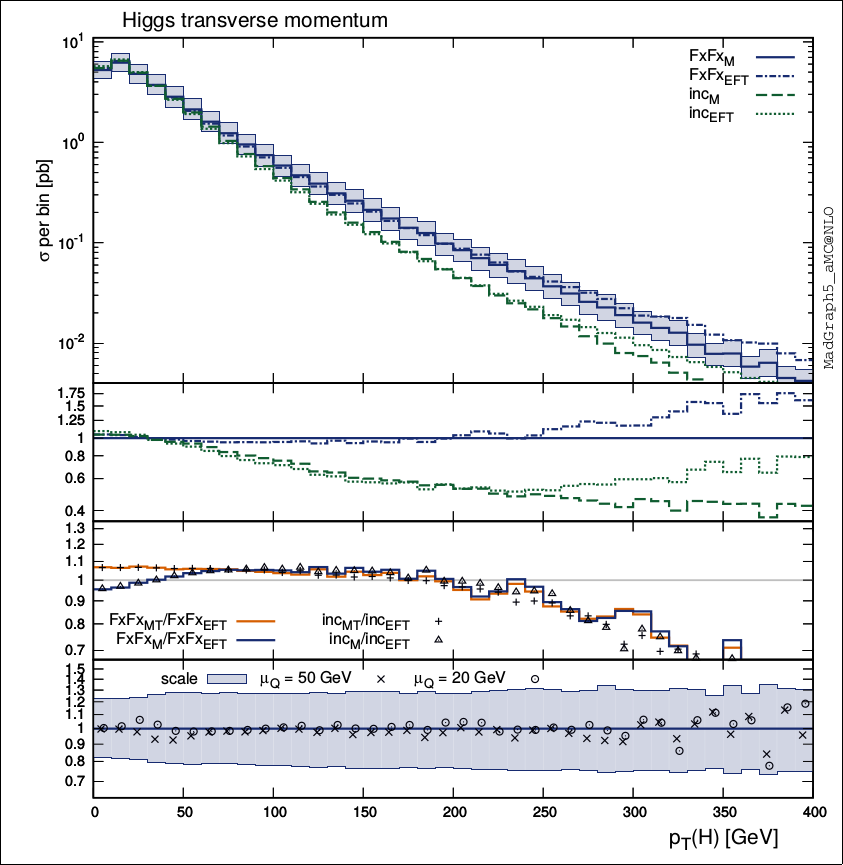}\hspace{1cm}
\includegraphics[height=2.2in,page=7]{final-all-yr4.pdf}
\caption{Left: inclusive $\pth$ distribution (plot taken from
  ref.~\cite{Frederix:2016cnl}). Effects due to heavy-quark mass
  effects and NLO+PS (FxFx) merging are visible. Right: the Higgs
  $\pth$ distribution as obtained with the NNLO+PS generator for
  $pp\to WH$ developed in ref.~\cite{Astill:2016hpa}. Details on cuts
  applied can be found in section I.5.5 of ref.~\cite{deFlorian:2016spz}.}
\label{fig:MC}
\end{figure}

As explained in section~\ref{sec:parton}, including exact top-mass
effects is important to model precisely the large-$\pth$ spectrum in
gluon-fusion production. This issue has also been addressed in the
context of NLO+PS multijet merging, where results have been obtained
using the {\tt MEPS@NLO}~\cite{Buschmann:2014sia} and {\tt
  FxFx}~\cite{Frederix:2016cnl} methods. In both cases, different
approximations were made to simulate the unknown mass effects in the
(2-loop) virtual corrections; notably, in~\cite{Frederix:2016cnl},
also $b$-quark mass effects have been considered. The left panel of
fig.~\ref{fig:MC} allows to appreciate the size of heavy-quark effects
as well as how important is it to merge (at NLO) the inclusive NLO+PS
prediction to higher jet multiplicities (in this case, up to 2 jets at
NLO).

Top-quark mass effects are also important in other contexts: for
instance, although it enters only at NNLO in the $pp\to ZH$ cross
section, the loop-induced $gg\to ZH$ matrix element is numerically
important, and gives sizeable effect in some phase-space corners. In
the context of multijet-merging, these effects were included by
merging (at LO) the $2\to 2$ ($gg\to ZH$) and $2\to 3$ ($gg\to ZH+1$
jet) loop-induced processes~\cite{Hespel:2015zea,Goncalves:2015mfa}.

Multi-jet merging is at the core of all the three methods currently
available to match NNLO computations with parton showers (NNLO+PS). So
far, this has been achieved only for color-singlet production, and for
relatively simple processes,~\emph{i.e.} for
Drell-Yan~\cite{Hoeche:2014aia,Karlberg:2014qua,Alioli:2015toa},
$gg\to H$ and $pp\to VH$ production. In particular, the ``improved
{\tt MiNLO}''~\cite{Hamilton:2012rf} and {\tt
  UNLOPS}~\cite{Lonnblad:2012ix} techniques are at the core of the
NNLO+PS accurate results obtained for gluon-fusion
production~\cite{Hamilton:2013fea,Hoche:2014dla}. Including mass
effects is obviously relevant also for NNLO+PS tools, and indeed, in
ref.~\cite{Hamilton:2015nsa}, a scheme to include top and bottom mass
effects in the {\tt MiNLO+POWHEG} result was proposed. It is
foreseable that results using the {\tt Geneva} approach will also
appear in the future.

The simulation of $VH$ production has recently witnessed several
results. On the one hand, {\tt MiNLO}-merged results for $WH$ and
$WH+1$ jet~\cite{Luisoni:2013kna} were upgraded to NNLOPS
accuracy~\cite{Astill:2016hpa} too, using the results
from~\cite{Ferrera:2013yga}: a representative plot showing the
difference between a NLO+PS merged result and the upgrade at NNLO for
a standard observable as $\pth$ is shown in the right-hand side panel
of fig~\ref{fig:MC}.  The inclusion of higher-order effects in the
simulation of the $H\to b\bar{b}$ decay will be very useful for some
searches aiming at measuring the bottom-quark Yukawa. Work is in
progress to include NLO corrections to this decay channel in
NNLO+PS-accurate event generator~\cite{Astill:tbp}, using the
techniques first developed to deal with NLO corrections to top-quark
decays~\cite{Campbell:2014kua,Jezo:2015aia}.

The combination of EW and QCD NLO corrections is a topic which is
receiving more and more attention. Although so far most of the studies
have focused on processes involving weak bosons in the final state, a
first result relevant for the modeling of Higgs signal was obtained
by the authors of ref.~\cite{Granata:2017iod}: fixed-order NLO QCD+EW calculations
were combined with a QCD+QED parton shower for the process $pp\to VH$
in association with 0 and 1 jet, by means of a combination of {\tt
  MiNLO} and the resonant-aware method~\cite{Jezo:2015aia}
in the {\tt POWHEG} framework.

To conclude this section, I'd like to mention a few publications which
might give an idea of where future developments could be expected in
the years to come. The authors of ref.~\cite{Frederix:2015fyz} have
proposed a method to lift one of the limitations of the {\tt MiNLO}
method, namely the fact that its NLO accuracy was proven only for the
two lowest jet multiplicities,~\emph{i.e.} the ``+0'' and ``+1'' jet
regions: it was indeed shown how to merge $H+2j$ at NLO together with
$H+j$ at NLO, as well as inclusive Higgs production at NNLO. So far
this is the most accurate result obtained with {\tt MiNLO} for the
$gg\to H$ process, and a possible future development could be to
include NNLO corrections for the $H+1$ jet region.

It is also clear that the limited logarithmic accuracy of parton
showers will soon become a bottleneck. Not surprisingly, in the last
couple of years many studies addressing this point were performed,
spanning from comprehensive estimates of uncertainties related to the
parton-showering stage of the Monte Carlo
simulation~\cite{Hoeche:2014lxa,Bellm:2016voq,Mrenna:2016sih} to first
attempts to improve accuracy of parton shower algorithms in
general~\cite{Nagy:2007ty}, and their logarithmic accuracy in
particular~\cite{Hartgring:2013jma,Li:2016yez,Hoche:2017hno}.




\Acknowledgements I am grateful to the conveners of the ``Higgs
physics in the Standard Model and beyond'' parallel session for the
invitation.

\end{document}

%% file: econfmacros.tex
\def\beq{\begin{equation}}
\def\eeq#1{\label{#1}\end{equation}}
\def\eeqn{\end{equation}}

\def\beqa{\begin{eqnarray}}
\def\eeqa#1{\label{#1}\end{eqnarray}}
\def\eeqan{\end{eqnarray}}

\let\bar=\overbar

\def\Dslash{\not{\hbox{\kern-4pt $D$}}}
\def\dslash{\not{\hbox{\kern-2pt $\del$}}}

\def\msb{{\bar{\ssstyle M \kern -1pt S}}}

